\documentclass[a4paper]{amsart}

\usepackage{amssymb,amsmath}
\newcommand{\ii}{\mathrm{i}}

\newcommand{\hh}{\mathcal{H}}

\newcommand{\rank}{\mathop{\mathrm{rank}}}
\begin{document}

\title[Noncommutative \& False Noncommutative]{A Note on Noncommutative and
 False Noncommutative spaces}%
\author{Corneliu Sochichiu}%
\address{Institutul de Fizic\u a Aplicat\u a
A\c S, str. Academiei, nr. 5, Chi\c sin\u au, MD2028
MOLDOVA}%
\address{Bogoliubov Laboratory of Theoretical Physics\\
Joint Institute for Nuclear Research\\ 141980 Dubna, Moscow Reg.\\
RUSSIA}
\email{sochichi@thsun1.jinr.ru}%

\thanks{Work supported by RFBR grant \#  99-01-00190, INTAS grant \# 950681,
and Scientific School support grant 96-15-0628}%

\begin{abstract}
We show that the algebra of functions on noncommutative space
allows two different representations. One is describing the
genuine noncommutative space, while another one can be rewritten
in commutative form by a redefinition of generators.
\end{abstract}
\maketitle

Noncommutative geometry (for a recent review see e.g.
\cite{Connes:2000by}) plays an important r\^{o}le in both string
theory, since it provides a tool for description of brane
dynamics. There is a wide activity and literature in this field,
so we refer reader to the appropriate reviews, e.g.
\cite{Seiberg:1999vs} and references therein/thereon.

The flat noncommutative space is given by coordinates $x^\mu$,
$\mu=1,\dots,D$, satisfying the noncommutativity relations,
\begin{equation}\label{commutator}
  [x^\mu,x^\nu]=\ii \theta^{\mu\nu},
\end{equation}
where for simplicity we assume that the matrix formed of
$\theta^{\mu\nu}$ is nondegenerate (therefore the space-time
dimensionality is even), the generalisation to a degenerate
$\theta^{\mu\nu}$ being also possible. These coordinate can be
viewed as generators of algebra of operators acting on a Hilbert
space $\hh$. The properties of the representation of algebra
(\ref{commutator}) are less discussed in the physicist's
literature, so we hope that this note may serve at least for the
pedagogical purposes. We are going to show that the above algebra
can both generate an irreducible representation which is the input
of a genuine noncommutative space or it be incorporated in a
larger algebra, in the last case the noncommutativity is fake and
can be eliminated by a simple (but nonlocal) shift of coordinates.

In addition to $x^\mu$ satisfying eq. (\ref{commutator}), consider
also the translation operators $p_\mu$, having canonical
commutators with $x^\mu$,
\begin{equation}\label{pmu}
  [p_\mu,x^\nu]=-\ii \delta_\mu^\nu.
\end{equation}

One can observe, that the quantity
$\pi_\mu=p_\mu+\theta^{-1}_{\mu\nu}x^\nu$, where
$\theta^{-1}_{\mu\nu}$ is the inverse matrix to $\theta_{\mu\nu}$,
commute with all the $x^\mu$,
\begin{equation}\label{zero}
  [\pi_\mu,x^\nu]=[p_\mu,x^\nu]+\theta^{-1}_{\mu\alpha}
  [x^\alpha,x^\nu]=0
\end{equation}

From now, there are (at least) two possibilities: depending on if
the representation of algebra generated by $x$'s alone is
irreducible or not on the Hilbert space $\hh$. In the first case,
one immediately has (by the virtue of the Schur's lemma), that
$\pi_\mu=0$. Thus, one has the translation operators $p_\mu$
expressed in terms of $x^\mu$, the only independent operators,
\begin{equation}\label{irrep}
  p_\mu=-\ii \theta^{-1}_{\mu\nu}x^\nu,
\end{equation}
which leads to the commutators,
\begin{equation}\label{[pp]}
  [p_\mu,p_\nu]=-\ii \theta^{-1}_{\mu\nu}.
\end{equation}

This is the standard $D$-dimensional noncommutative plane. In this
case one can pass through the Lorentz transformation to the
coordinates $z^m$ and $\bar{z}_{\bar{m}}$, satisfying,
\begin{equation}\label{zz}
  [\bar{z}_m,z_n]=-\ii \theta^{(m)}\delta_{m n},
\end{equation}
while the matrix $\theta^{\mu\nu}$ is brought to the block
diagonal form with $D/2$ antisymmetric $(2\times 2)$-blocks,
$m=1,\dots,D/2$,
\begin{equation*}
  \begin{pmatrix}
  0&\theta_{(m)}\\
  -\theta_{(m)}&0
  \end{pmatrix}
\end{equation*}
There no sum over $m$ in eq. (\ref{zz}) is assumed.

This is represented as the \emph{standard Heisenberg algebra in
$D/2$ dimensions} with planckian constant depending on the
direction.

Consider now a different situation. Let this turn $x^\mu$ not to
generate an irreducible representation by themselves, but only
together with $p_\mu$.

In this case, in spite of commuting with all $x^\mu$, the quantity
$\pi_\mu$, should not vanish unless it commutes also with all
$p_\mu$. The last happens when the commutators of $p_\mu$ satisfy
eq. (\ref{[pp]}). In this case, all $\pi_\mu=0$, and one returns
back to the previous situation. Consider the case when $p_\mu$ has
a generic $c$-number commutator,
\begin{equation}\label{B}
  [p_\mu,p_\nu]=-\ii B_{\mu\nu},
\end{equation}
where $B_{\mu\nu}$ is an antisymmetric matrix.

It is not difficult to see, that the criterium for the existence
of a nontrivial $\pi_\mu$ is,
\begin{equation}\label{rank}
  r\equiv\rank (B_{\mu\nu}-\theta^{-1}_{\mu\nu})>0,
\end{equation}
$r$ giving the number of independent operators $p_\mu$, the
maximal case, when $p_\mu$ are all independent, being when
$\det(B-\theta^{-1})\neq 0$. In this case, one has an usual
\emph{$D$-dimensional Heisenberg algebra} generated by $p_\mu$ and
$q^\mu$ (compare with $D/2$-dimensional one in the previous case).
One can again pass to the canonical variables, $P_\mu$ and $X^\mu$
which have the standard commutators,
\begin{equation}\label{stand}
  [P_\mu,X^\nu]=-\ii\delta_\mu^\nu,\qquad [P,P]=[X,X]=0,
\end{equation}
by a shift,
\begin{align}\label{shift-x}
  &x^\mu\rightarrow X^\mu=x^\mu+\xi^{\mu\alpha}p_\alpha, \\
  \label{shift-p}
  &p_\mu\rightarrow P_\mu=p_\mu+\zeta_{\mu\alpha}x^\alpha,
\end{align}
where $\xi^{\mu\nu}$ and $\zeta_{\mu\nu}$ satisfy the following
equations,
\begin{align}\label{unu}
  &\theta^{\mu\nu}-2\xi^{[\mu\nu]}-\xi^{\mu\alpha}B_{\alpha\beta}
  \xi^{\nu\beta}=0, \\ \label{doi}
  &-B_{\mu\nu}+2\zeta_{[\mu\nu]}+\zeta_{\mu\alpha}\theta^{\alpha\beta}
  \zeta_{\nu\beta}=0,\\ \label{trei}
  &\xi^{\mu\alpha}B_{\alpha\nu}+\zeta_{\mu\alpha}\theta^{\alpha\nu}
  +\zeta_{\mu\alpha}\xi^{\nu\alpha}=0,
\end{align}
``$[\dots]$'' denotes the antisymmetric part of the matrix. In
what follows we will not try to analyse the solutions to eqs.
(\ref{unu}-\ref{trei}), but only point that the existence of
solution follows from the Darboux theorem. In the case when
$B_{\mu\nu}=0$, the above equations greatly simplify and one has,
\begin{align}\label{b=0}
  &X^\mu=x^\mu+\frac{1}{2}\theta^{\mu\nu}p_\nu, \\
  &P_\mu=p_\mu.
\end{align}

As a result, we have a \emph{commutative} space generated by
$X^\mu$ common with vector algebra on it generated by $P_\mu$.
However, from the point of view of the Quantum Field Theory the
shift $x\rightarrow X$ and $p\rightarrow P$ produce a nonlocal
field redefinitions containing an infinite number of derivatives.

Finally, let us note that the last situation is common (up to interchange
$p\leftrightarrow x$) in usual mechanics. In the presence of a constant
electromagnetic field the translation operators $p_\mu$ have a nontrivial
commutator, like in eq. (\ref{B}), where $B_{\mu\nu}$ is the field strength. This,
however, does not lead to any noncommutativity of the space-time.

A more appropriate example is given by a model with higher derivative terms
considered in \cite{Lukierski:1997br}, where it was shown that a noncommutativity
removable by transformation similar to (\ref{b=0}) arises.

\subsection*{Discussions.} Let us summarise the results so far obtained. We
considered the algebra of operators $x^\mu$ and $p_\mu$ satisfying
commutator relations (\ref{commutator},\ref{pmu}) and (\ref{B}).
We required this algebra to have irreducible representation on the
Hilbert space $\hh$. We obtained that in the case when
$B=\theta^{-1}$ this is representation of the $D/2$-dimensional
Heisenberg algebra and this is ``genuine'' noncommutative space,
while in the case when $\det(B-\theta)\neq 0$ it is one of
$D$-dimensional Heisenberg algebra and it is reduced, although by
nonlocal transformation, to the usual commutative space.

Using the equivalence of Heisenberg algebra representations in
different dimensions \cite{Sochichiu:2000bg}, which falls in
general context of background independence of noncommutative
theories \cite{Seiberg:2000zk}, one may conclude that gauge models
based on these representations are also equivalent. We think that
this equivalence should be provided by some analog of the
Seiberg--Witten map \cite{Seiberg:1999vs} when the
noncommutativity is treated dynamically as in
\cite{Harvey:2000jt,Sochichiu:2000rm,Gopakumar:2000xx}.

\subsection*{Acknowledgements.} I benefitted from useful
discussions with A.A.~Slavnov and F.~Toppan.
\providecommand{\href}[2]{#2}\begingroup\raggedright\endgroup

\end{document}